\documentclass[final]{m-ws-ijmpcs}
\pdfoutput=1

\begin{document}

\markboth{A.~A.~Chernitskii}{Induced gravitation in nonlinear field models}

%
\catchline{9th A. Friedmann International Seminar and}{3rd Casimir Symposium 2015}{Vol. 41 (2016) 1660119 (7 pages)}{DOI: 10.1142/S2010194516601198}{}
%

\title{Induced gravitation in nonlinear field models}

\author{Alexander A. Chernitskii}

\address{Department of Mathematics\\ St. Petersburg State Chemical Pharmaceutical Academy\\Prof. Popov str. 14, St. Petersburg, 197022, Russia}
\address{A. Friedmann Laboratory for Theoretical Physics\\St. Petersburg, Russia\\AAChernitskii@mail.ru}

\maketitle

\begin{history}

\received{19 September 2015}
\published{18 March 2016}
\end{history}
\begin{abstract}
The description of gravitation in the framework of soliton interaction is considered for two nonlinear field models.
These models are Born -- Infeld nonlinear electrodynamics and so-called Born -- Infeld type scalar field model.
The last model can also be called the extremal space-time film one because of the specific form of the appropriate variational principle.
Gravitational interaction is considered in the context of unification for all interactions of material particles.
It is shown that long-range interaction of solitons of the models appears as force one and metrical one.
The force interaction can be interpreted as electromagnetic one. The metrical interaction can be interpreted as gravitational one.

\end{abstract}

\ccode{PACS numbers: 04.50.-h;12.10.-g}

\def\df{\mathrm{d}}
\def\bfgr#1{\pmb{#1}}
\def\p{\partial}
\def\dfrac#1#2{{\displaystyle\frac{#1}{#2}}}
\def\stTD#1#2{\hbox to 0em{\mathsurround=0em $\stackrel{#1}{\makebox[0pt]{} #2}$\hss} \phantom{#2}}\def\stscript#1#2{\hbox to 0em{\mathsurround=0em ${\scriptstyle\stackrel{#1}{\makebox[0pt]{} #2}}$\hss} \phantom{#2}}\def\stscriptscript#1#2{\hbox to 0em{\mathsurround=0em ${\scriptscriptstyle\stackrel{#1}{\makebox[0pt]{} #2}}$\hss} \phantom{#2}}\def\st#1#2{\mathchoice{\stTD{#1}{#2}}{\stTD{#1}{#2}}{\stscript{#1}{#2}}{\stscriptscript{#1}{#2}}}
\def\comb#1#2#3{{\mathsurround 0pt\hbox to 0pt {\hspace*{#3}\raisebox{#2}{${#1}$}\hss}}}
\def\combs#1#2#3{{\mathsurround 0pt\hbox to 0pt {\hspace*{#3}\raisebox{#2}{${\scriptstyle #1}$}\hss}}}
\def\combss#1#2#3{{\mathsurround 0pt\hbox to 0pt {\hspace*{#3}\raisebox{#2}{${\scriptscriptstyle #1}$}\hss}}}

\def\bFem{\mathbf{F}}
\def\Fem{F}
\def\bfem{\mathbf{G}}
\def\fem{G}
\def\him{\epsilon}
\def\Sur{\Sigma}
\def\dSur{{\rm d}\hspace{-0.3ex}\Sigma}
\def\Act{\mathcal{A}}
\def\invEBa{\combs{-}{0.4ex}{0.15ex}\mathcal{I}}
\def\invEBb{\combs{-}{0.3ex}{0.55ex}\mathcal{J}}
\def\xc{\bar{x}}
\def\metrEff{\mathchoice{\combs{\sim}{1ex}{0.2ex}\mathfrak{m}}{\combs{\sim}{1ex}{0.2ex}\mathfrak{m}}{\combss{\sim}{0.66ex}{0.05ex}\mathfrak{m}}{}{}}
\def\metr{\mathfrak{m}}
\def\metrp{\mathchoice{\comb{-}{-0.9ex}{0ex}\mathfrak{m}}{\comb{-}{-0.9ex}{0ex}\mathfrak{m}}{\combs{-}{-0.75ex}{-0.1ex}\mathfrak{m}}{}{}}
\def\Vol{\overline{V}}
\def\dVol{\df\mspace{-2mu}\Vol}
\def\Vols{V}
\def\dVols{\df\mspace{-2mu}\Vols}
\def\Surc{\sigma}
\def\dSurc{{\rm d}\hspace{-0.3ex}\sigma}
\def\ffun{\Phi}
\def\xxx{\chi}
\def\Ae{A}
\def\eqdef{\doteqdot}
\def\Fo{\mathbb{F}}
\def\xp#1{\comb{\cdot}{-0.9ex}{0.3ex}{#1}}
\def\EMT{T}
\def\EMTc{\mathchoice{\combs{-}{1.5ex}{0.2ex}{\EMT}}{\combs{-}{1.5ex}{0.2ex}{\EMT}}{\combss{-}{1.1ex}{0.05ex}{\EMT}}{\combss{-}{1.1ex}{0.05ex}{\EMT}}}
\def\EMTi{\mathchoice{\combss{\infty}{1.8ex}{0.15ex}\EMT}{\combss{\infty}{1.85ex}{0.15ex}\EMT}{\combss{\infty}{1.25ex}{-0.12ex}\EMT}{\combss{\infty}{1.2ex}{-0.12ex}\EMT}}
\def\AMT{\mathchoice{\combs{\circ}{0.9ex}{0.9ex}{M}}{\combs{\circ}{0.9ex}{0.9ex}{M}}{\combss{\circ}{0.7ex}{0.65ex}{M}}{\combss{\circ}{0.7ex}{0.65ex}{M}}}
\def\AMTc{\mathchoice{\combs{-}{0ex}{0.41ex}{\AMT}}{\combs{-}{0ex}{0.42ex}{\AMT}}{\combss{-}{-0.05ex}{0.25ex}{\AMT}}{\combss{-}{-0.05ex}{0.25ex}{\AMT}}}
\def\x{\mathtt{x}}
\def\k{\mathtt{k}}
\def\Energy{\mathbb{E}}
\def\bEMV{\pmb{\mathbb{P}}}
\def\EMV{\mathbb{P}}
\def\bAMV{\pmb{\mathbb{J}}}
\def\AMV{\mathbb{J}}

\section{Introduction}
\label{introd}

The description of gravitational interaction is possible in the framework of
nonlinear field models, where the  physical particles are represented by its soliton solutions.
In this consideration all interactions between the particles are caused by nonlinearity of the model.

Soliton solution or soliton is a space-localized solution of nonlinear field model.
A distinguishing characteristic of nonlinear field equations is that the sum of solutions is not solution.
That is the superposition principle, which is characteristic for linear equations, here does not work.

The violation of superposition principle must be considered as an interaction of soliton-particles.
We can consider the sum of soliton solutions as an initial approximation for an appropriate multisoliton solution.
Then we must make an iterative process to obtain the solution.

 A modification of soliton trajectories, which we will have here, must be interpreted as the interaction between the soliton-particles.

If distance between solitons in the initial approximation sufficiently great, we have a long-range
soliton interaction.

Thus
in the framework of this approach we consider an unification of two known long-range interactions between physical particles
that are electromagnetic and gravitational ones.

The idea of unification for gravitation and electromagnetism in the framework of soliton interaction
was proposed in works by
author,\cite{Chernitskii1992}\cdash\cite{Chernitskii2004a}
where nonlinear electrodynamics models was considered.

In this consideration the electromagnetic interaction appears as soliton interaction of the first order by a small field of distant solitons.
In this case we have the Lorentz force from the integral conservation law for energy-momentum.
This interaction can also be called the force one. Here it will be considered briefly in the appropriate section.

The gravitational interaction appears as soliton interaction of the second order by the small field
of distant solitons.
In this case we have an effective Riemann space for soliton motion, which is caused by the field of distant solitons.

The description of gravitation with the effective Riemann space in soliton dynamics was proposed in
works by author.\cite{Chernitskii1992,Chernitskii1999,Chernitskii1998b}
Here we can talk on induced gravitation because the gravitational interaction appears in
models for fields which is not customary gravitational ones.
This interaction can also be called the metrical one. Here it will be considered in the appropriate section.

Mathematical base for the effective Riemann space in this consideration is the specific form of characteristic equation for several field models.
The appropriate characteristic equation for Born -- Infeld nonlinear
electrodynamics was considered in several works.\cite{Chernitskii1998b}\cdash\cite{Novello2000}

In the present work we consider a scalar field model in comparison with nonlinear electrodynamics one
for the description of long-range interactions of physical particles.

\section{Extremal World Volume}
\label{extrv}
Let us consider the action which has the following world volume form:
 \begin{equation}
\label{35135655}
\Act  =\int\limits_{\Vol}\!\sqrt{|\mathfrak{M}|}\;(\mathrm{d}x)^{4}
\;,
\end{equation}
where  $\Vol$ is a space-time volume,
$\mathfrak{M} \eqdef \det(\mathfrak{M}_{\mu\nu})$ is determinant of tensor field
$\mathfrak{M}_{\mu\nu} = \mathfrak{M}_{\mu\nu} (\x)$,
$(\mathrm{d}x)^{4} \eqdef \mathrm{d}x^0\mathrm{d}x^1\mathrm{d}x^2\mathrm{d}x^3$.

The tensor $\mathfrak{M}_{\mu\nu}$ can also be called the world tensor.

We have the usual action principle
which is the following necessary condition for extremal world volume:
\begin{equation}
\label{VarPr}
\delta\Act = 0
\;.
\end{equation}

\begin{subequations}\label{391957151}
Here we consider the following two types of the world tensor and the appropriate field models.
\begin{itemize}
\item For four-vector field $\Ae_\mu (\x)$ we can write the world tensor which gives the known Born -- Infeld nonlinear electrodynamics:
\begin{equation}
\label{540550431}
 \mathfrak{M}_{\mu\nu} = \metr_{\mu\nu} + \chi^2\,\Fem_{\mu\nu}
\;,\quad
\Fem_{\mu\nu} \eqdef \frac{\p \Ae_\nu}{\p x^\mu} {}-{}
\frac{\p \Ae_\mu}{\p  x^\nu}
\;.
\end{equation}
\item  For scalar field $\Phi_{\mu} (\x)$ we can write the world tensor which is appropriate to
a relativistic generalization for the model of minimal two-dimensional surface in three-dimensional space.
Thus we consider a four-dimensional surface or film in five-dimensional space-time
which can be called also the space-time film.
Sometimes the appropriate field model is called the scalar Born -- Infeld one.
For this case we have
\begin{equation}
\label{547036541}
  \mathfrak{M}_{\mu\nu} = \metr_{\mu\nu} + \chi^2\,\Phi_{\mu}\,\Phi_{\nu}
  \;,\quad
\Phi_{\mu} \eqdef \frac{\p \Phi}{\p x^\mu}
\;.
\end{equation}
\end{itemize}
\end{subequations}

Here in (\ref{391957151}) $\metr_{\mu\nu}$ are the components of metric tensor for flat space-time, $\chi$ is
a dimensional constant.

\begin{subequations}\label{392107891}
We can simplify the expression for the world tensor determinant.
\begin{itemize}
\item For Born -- Infeld nonlinear electrodynamics we have
\begin{align}
\nonumber
 \mathfrak{M}  &= \metr \left(1 {}-{} \chi^2\,\invEBa {}-{} \chi^4\,\invEBb^2\right)
\;,\\
\label{579195702}
&
 \invEBa \eqdef
 -\frac{1}{2}\,\metr^{\mu\sigma}\,\metr^{\nu\rho}\,
\Fem_{\mu\nu}\,\Fem_{\sigma\rho}
\;,\quad
\invEBb \eqdef
-
\frac{1}{8}\,\him^{\mu\nu\sigma\rho}\, \Fem_{\mu\nu}\,\Fem_{\sigma\rho}
\;.
\end{align}
\item For space-time film we have
\begin{equation}
\label{570904461}
 \mathfrak{M}  = \metr \left(1 {}+{} \chi^2\,\metr^{\mu\nu}\,\Phi_{\mu}\,\Phi_{\nu}\right)
\;.
\end{equation}
\end{itemize}
\end{subequations}

\section{Energy-Momentum Tensor}
\label{enmomt}
Let us write symmetrical energy-momentum tensors for both models
in Cartesian coordinates with Minkowski metric $\metrp^{\mu\nu}$.
 At first we define the energy-momentum tensors in the form which is obtained from the world volume
 (\ref{35135655}) directly.
\begin{subequations}\label{669566201}
\begin{itemize}
\item For nonlinear electrodynamics the symmetrical energy-momentum tensor is
\begin{align}
\nonumber
\EMTc^{\mu\nu}
&= \frac{1}{4\pi}\,\left(
  \fem^{\mu\rho}\,\Fem^{\nu}_{.\rho}
  {}-{} \frac{1}{\chi^2}\,\metrp^{\mu\nu}\,\sqrt{\left|\mathfrak{M}\right|} \right)
  \;,\\
\label{Expr:f}
&\fem^{\mu\nu} \eqdef
\frac{1}{\sqrt{\left|\mathfrak{M}\right|}}\,\left(\Fem^{\mu\nu}  {}+{}
\frac{\chi^2}{2}\,\invEBb\,\him^{\mu\nu\sigma\rho}\,
\Fem_{\sigma\rho}\right)
\;.
\end{align}
\item For space-time film the canonical energy-momentum tensor is symmetrical. It is
\begin{equation}
\label{595137671}
\EMTc^{\mu\nu} = \frac{1}{4\pi}\left(\frac{\Phi^{\mu}\,\Phi^{\nu}}{\sqrt{\left|\mathfrak{M}\right|}} - \frac{\metrp^{\mu\nu}}{\chi^2}\,\sqrt{\left|\mathfrak{M}\right|}\right)
\;.
\end{equation}
\end{itemize}
\end{subequations}

The obtained tensors will be used below. But these tensors do not equal zero when the field is vanished.

To have finite full energy for solutions which are decreasing at infinity, we consider also the following regularized
energy-momentum tensor:
\begin{equation}
\label{613171741}
 \EMT^{\mu\nu} = \EMTc^{\mu\nu} - \EMTi^{\mu\nu}
\;,
\end{equation}
where $\EMTi^{\mu\nu}$ is regularizing tensor.

In most cases we can use the following regularizing energy-momentum tensor:
\begin{equation}
\label{719207201}
\EMTi^{\mu\nu} =
-\frac{1}{4\pi\,\chi^2}\,\metrp^{\mu\nu}
\;.
\end{equation}

Both tensors $\EMT^{\mu\nu}$ and $\bar{\EMT}^{\mu\nu}$ satisfy the differential conservation law
\begin{equation}
\label{770165531}
\frac{\p \EMTc^{\mu\nu}}{\p x^{\mu}}  = \frac{\p \EMT^{\mu\nu}}{\p x^{\mu}} = 0
\;.
\end{equation}

\section{Field Equations and Effective Metrics}
\label{feandco}
\begin{subequations}\label{784691541}
 It is notable that the field equations for the models under consideration can be written in the following forms
 in Cartesian coordinates.
 \begin{itemize}
 \item For nonlinear electrodynamics we have
\begin{equation}
\label{Eq:EulerBIA1}
\left(
\metrEff^{\mu\sigma}\,\metrEff^{\nu\rho}
{}-{}
\metrEff^{\mu\rho}\,\metrEff^{\nu\sigma}
\right)
\frac{\p^2 \Ae_\rho}{\p x^\nu \p x^\sigma}
{}={} 0
\;.
\end{equation}
 \item For space-time film we have
 \begin{equation}
\label{4039893311}
\metrEff^{\mu\nu}\,\frac{\p^{2}\,\ffun}{\p x^{\mu}\,\p x^{\nu}}  = 0
\;.
\end{equation}
 \end{itemize}
 \end{subequations}
 Here in (\ref{784691541}) we introduce the designation
 \begin{equation}
\label{588743701}
\metrEff^{\mu\nu}  \eqdef -4\pi\,\chi^2\,\EMTc^{\mu\nu}
\;,
\end{equation}
where the energy-momentum tensor $\EMTc^{\mu\nu}$ is defined in (\ref{669566201}) for both models.

As we can see the model equations (\ref{784691541}) become suitable linear ones with the substitution
\begin{equation}
\label{793123221}
\metrEff^{\mu\nu}  \to \metrp^{\mu\nu}
\;.
\end{equation}
In this case we obtain the appropriate relativistically invariant linear equations.

Thus the tensor $\metrEff^{\mu\nu}$ can be called the effective metrics.
This appellation is justified also with the consideration of a model characteristic equation
and the appropriate soliton interaction. This theme is discussed in section \ref{metrint}.

\section{Force Interaction and Electromagnetism}
\label{forceint}
  Let us consider a constant three-dimensional volume $\Vols$ with the border $\Surc$.

  We define the momentum four-vector of field in the volume $\Vols$ as
 \begin{equation}
\label{23695072}
\EMV^{\mu}_{\Vols}\doteqdot \int\limits_{\Vols}\EMT^{\mu 0}\,\dVols
\;.
 \end{equation}

 Also let us define the force of the field on the two-dimensional surface $\Surc$ as
 \begin{equation}
 \label{294678411}
\Fo^\mu_{\Surc} \eqdef -\int\limits_{\Surc} \EMT^{\mu i}\,\dSurc_i
\;,
 \end{equation}
 where $\dSurc_i$ are components of the outer surface element vector.

 Integration by parts of the energy-momentum differential conservation law (\ref{770165531}) for tensor $\EMT^{\mu\nu}$  gives
 \begin{equation}
 \label{478204531}
\int\limits_{\Vol}
\frac{\partial \EMT^{\mu\nu}}{\partial x^\nu}\;\dVol = 0
\quad\Longrightarrow\qquad\int\limits_{\Sur}\EMT^{\mu\nu}\,\dSur_\nu = 0\;,
 \end{equation}
where $\Vol$ is four-dimensional volume including the three-dimensional volume $\Vols$ and time interval
$\left[\xc^0-{\Delta x^0}/{2}\,,\;\xc^0+{\Delta x^0}/{2}\right]$,
$\Sur$ is its three-dimensional boundary hypersurface,
$\dSur_\nu$ are components of the outer hypersurface element four-vector.

Using introduced designations for momentum and force we have
 \begin{equation}
 \label{480922341}
\Delta\EMV^\mu_{\Vols} = \Delta\Fo^\mu_{\Surc}
\;,
 \end{equation}
where
\begin{subequations}
 \begin{align}
 \label{496313401}
\Delta\EMV^\mu_{\Vols}
&\eqdef \int\limits_{\Sur} \EMT^{\mu 0}\,\dSur_0
= \EMV^\mu_{\Vols}
\biggr|_{x^0 = \xc^0 {}-{} \frac{\Delta x^0}{2}}^{x^0 = \xc^0 {}+{} \frac{\Delta x^0}{2}}
\;,\\
\Delta\Fo^\mu_{\Surc}
&\eqdef -\int\limits_{\Sur} \EMT^{\mu i}\,\dSur_i
= \int\limits_{\xc^0 {}-{} \frac{\Delta x^0}{2}}^{\xc^0 {}+{} \frac{\Delta x^0}{2}} \Fo^{\mu}_{\Surc}\,
\df x^0
\;.
 \end{align}
\end{subequations}

For $\Delta x^0 \to 0$ we have the following integral dynamical law for the field in the three-dimensional volume $\Vols$:
\begin{equation}
\label{509874761}
\dfrac{\df \EMV^\mu_{\Vols}}{\df \xc^0} = \Fo^\mu_{\Surc}
\;.
\end{equation}

Application of the obtained dynamic law to the soliton interaction gives the appropriate soliton dynamic equation in some approximation.
In this case the three-dimensional volume $\Vols$ is the localization region of the soliton.
The force $\Fo^\mu_{\Surc}$ is caused by a small field of the distant solitons.
Thus this soliton interaction can be called the force interaction of solitons.

It must be emphasized that the obtained dynamical law (\ref{509874761}) is model independent.
Also its relativistic invariance is evident for the models under consideration.

The considered force interaction applying to Born -- Infeld electrodynamics gives the electromagnetic
 interaction between soliton-particles with the Lorentz force.\cite{Chernitskii1999}

 But the second model under consideration, that is the model of space-time film, gives also this soliton interaction.
This is supported by universality of the considered force interaction.
In this case the electromagnetic force of distant solitons appears as essentially nonlocal characteristic
  of interaction, which is caused by nonlocal definition of the force.

Thus the force interaction considered here is electromagnetic one.

\section{Metrical Interaction and Gravitation}
\label{metrint}
Both models under consideration have the following notable characteristic
equation:\footnote{For Born -- Infeld electrodynamics see
Refs.~\refcite{Chernitskii1999}, \refcite{Chernitskii1998b} and \refcite{Chernitskii2012be}.}
\begin{equation}
\label{582613611}
\metrEff^{\mu\nu}\,k_\mu\,k_\nu = 0
\;,\quad
k_\mu {}\eqdef{} \frac{\partial {\cal S}}{\partial x^\mu}
\;,
\end{equation}
where ${\cal S} = {\cal S}(\{x^{\mu}\})$,
equation
\begin{equation}
\label{322256801}
{\cal S}= 0
\end{equation}
 gives a three-dimensional characteristic hypersurface of field model in four-dimensional space-time,
$\metrEff^{\mu\nu}$ is effective Riemann metrics which is defined in (\ref{588743701}).

Let us consider a soliton having a fast-oscillating part.
In this case the soliton is connected with the appropriate wave for which we can define the wave vector $k_{\mu}$.

Thus according to relativistic invariance of the model we have the following dispersion relation for such solitons:
\begin{equation}
\label{731994201}
\left|\metr^{\mu\nu}\,k_{\mu}\,k_{\nu}\right|  =  \xp{\omega}^2
\;,
\end{equation}
where
$\metr^{\mu\nu}$ is the coordinate system metric,
$\xp{\omega}$ is a frequency of the soliton fast-oscillating part in its intrinsic coordinate system.

The case $\xp{\omega} \neq 0$ corresponds to a massive particle.
The case $\xp{\omega} = 0$ corresponds to a massless particle, specifically to photon.

It can be shown,\cite{Chernitskii1999,Chernitskii2012be}
that in the case of soliton long-range interaction, the soliton dispersion relation
(\ref{731994201})
is modified to
\begin{equation}
\label{396089081}
\left|\metrEff^{\mu\nu}\,k_{\mu}\,k_{\nu}\right|  =  \xp{\omega}^2
\;,
\end{equation}
where the effective metric $\metrEff^{\mu\nu}$ is caused by a field of distant solitons.

In this case the trajectory equation of the interacting soliton is the same that we have in general relativity theory, that is the
appropriate geodesic equation.

The type of soliton long-range interaction caused by the effective metric can be called the metrical one.
Because of the specific type of interacting soliton trajectory equation for this case, we can consider the accordance
of this interaction with gravitation.
It should be noted that in this case the wave properties of the interacting soliton is essential.

It can be shown,\cite{Chernitskii2012be,Chernitskii2002b}
to consider the metrical soliton interaction as the real gravitation we must take into account the wave background
field of all soliton-particles in the universe.

In this case we can have the Newtonian gravitation in the appropriate limit.
Also to have the real gravitation we must take into account the fast-oscillating part of distant solitons.

Thus the wave or quantum properties of soliton-particles is essential for the gravitation interpretation of
metrical interaction.
Thus in this approach the gravitation is essentially quantum.

\section{Conclusions}
\label{concl}

\begin{itemize}
\item  Two nonlinear field models with world volume type action are considered.
\begin{itemize}
\item The first model is the Born -- Infeld nonlinear electrodynamics.
\item The second one can be called the model of extremal four-dimensional film. The appropriate equation sometimes is called the scalar Born -- Infeld one.
\end{itemize}
\item Both models provide the possibility for
unification of gravitational and electromagnetic interactions of particles in the framework of interaction for solitons.
\begin{itemize}
\item The force interaction is obtained from the integral conservation law for momentum. This interaction can be interpreted as electromagnetic one.
\item The metrical interaction is obtained from the model characteristic equation. This interaction can be interpreted as gravitational one.
In this case we have the induced gravitation.
\end{itemize}
\end{itemize}

\end{document}